\documentclass[final,cmfonts]{aipproc}

\layoutstyle{6x9}

\begin{document}

\title{A Peculiar Dynamically Warped Theory Space}


\author{Christopher D. Carone}{
  address={Particle Theory Group, Department of Physics,
College of William and Mary, Williamsburg, VA 23187-8795}
}

\begin{abstract}
We study a supersymmetric deconstructed gauge theory in which a warp factor emerges dynamically,
driven by Fayet-Iliopoulos terms. The model is peculiar in that it possesses a global supersymmetry that
remains unbroken despite nonvanishing D-term vacuum expectation values.  Inclusion of gravity and/or
additional messenger fields leads to the collective breaking of supersymmetry and to an unusual
phenomenology.
\end{abstract}

\maketitle


\section{Introduction}

In this talk\footnote{Presented at SUSY06, the 14th International Conference on Supersymmetry and the
Unification of Fundamental Interactions, Irvine, California, USA 12-17 June 2006. WM-06-109.},
we consider a four-dimensional, linear ``moose" model that deconstructs~\cite{decon} a slice of five-dimensional
Anti-de Sitter (AdS) space~\cite{deconads}.  The profile of link field vacuum expectation values (vevs)
along the moose can be chosen to replicate the effects of a warp factor in the higher-dimensional theory.
The question we study is whether the necessary profile can be generated dynamically and naturally.  In
Ref.~\cite{thepaper}, we present a number of examples where a monotonically varying warp factor is
obtained by assuming a translational symmetry along the moose and specific choices for boundary conditions
at its ends.  Here, we focus on a supersymmetric example in which a warp factor is driven by
Fayet-Iliopoulos  (FI) terms.  The model is peculiar in that it possesses a global supersymmetry that
remains unbroken despite nonvanishing D-term vevs.  Inclusion of gravity and/or additional messenger fields
leads to the collective breaking of supersymmetry, with interesting consequences.  Other applications
of deconstruction in model building can be found in Refs.~\cite{more1,more2}.

\section{The Model}

The model we consider is a 4D ${\cal N}$=1 SUSY  U(1)$^n$ moose theory.   The link fields consist of chiral
multiplets $\phi_i$ with charges $(q_i,q_{i+1})=(+1,-1)$, where $i$ labels the gauge group factor.
Conjugate superfields $\overline{\phi}_i$ are included to cancel anomalies.

The scalar potential for the link fields is given by
\begin{equation}
V_D=\sum_{i=1}^n D_i^2,
\label{eq:pot}
\end{equation}
where
\begin{equation}
D_i = g\left( |\phi_{i}|^2 - |\phi_{i-1}|^2 -
|\overline{\phi}_{i}|^2 + |\overline{\phi}_{i-1}|^2 + \xi_{i}\right) \,\,\,,
\label{eq:D-terms}
\end{equation}
and where we define $\phi_0=\phi_{n}=0$.  Here $g$ is the common gauge coupling and $\xi_i$
is the FI term for the $i^{th}$ group.  This potential is minimized when
\begin{equation}
\left<\phi_i\right>\,\left(\left<D_i\right>-\left<D_{i+1}\right>\right)=0\ ,
\end{equation}
which implies that the vacua of interest generically have equal $D$-terms,
\begin{equation}
\left<D_i\right>=\frac{\sum_j g\xi_j}{n}\equiv D \,\,.
\label{eq:dterms}
\end{equation}
The scalar vevs $v_i$ and $\overline{v}_i$ satisfy the recursion relation
\begin{equation}
(|v_{i+2}|^2-|\overline{v}_{i+2}|^2) - 2(|v_{i+1}|^2-|\overline{v}_{i+1}|^2)+
(|v_i|^2 - |\overline{v}_i|^2) = (\xi_{i+1}-\xi_{i+2}) \,\,\, ,
\end{equation}
which is a discretized form of
\begin{equation}
\frac{\partial^2|\phi(y)|^2}{\partial y^2} = -\frac{\xi'(y)}{a},
\end{equation}
where $a=1/(g v_1)$ is the lattice spacing.  Integrating this result and expressing the link
profile $\phi(y)$ in terms of the warp factor, one finds
\begin{equation}
\frac{\partial e^{-{f}(y)}}{\partial y}=(-g^2 \xi(y)+gD)a\,,\,\, \,\,\,\mbox{ where }
\,\,\,\,D/g=\int_0^R dy \,\xi(y)/R \,.
\end{equation}
Notice that any desired warp factor can be obtained by setting
\begin{equation}
\xi(y)=\widetilde{\xi}(y)+D/g = \widetilde{\xi}(y) +\int_0^R dy\, \xi(y)/R
\label{eq:desire}
\end{equation}
and choosing an appropriate function $\widetilde{\xi}(y)$.  However, Eq.~(\ref{eq:desire}) is
self-consistent only if
\begin{equation}
\int_0^R dy\,\widetilde{\xi}(y)=0  \,\,\,.
\end{equation}
A monotonically varying warp factor is possible provided that $\widetilde{\xi}(y)$ receives
an opposite sign contribution at the boundary.

\section{The Spectrum}

The Kaluza-Klein spectrum of this model is surprising, given the non-vanishing $\langle D_i \rangle$ in
Eq.~(\ref{eq:dterms}).  The masses of the vector and chiral multiplets originate from the kinetic terms
\begin{equation}
{\cal L}\supset\int d^4\theta\,
\sum_i \Phi_i^\dagger\,\exp\left[g(V_{i}-V_{i+1})\right]\,\Phi_i +
\overline{\Phi}_i^\dagger\, \exp\left[g(-V_{i}+V_{i+1})\right]\,\overline{\Phi}_i.
\end{equation}
One finds that the gauge boson mass matrix is
\begin{equation}
m_{\rm gauge}^2 = 2 g^2 \left(\begin{array}{ccccccc}
v_1^2 && -v_1^2 && &&  \\
-v_1^2 && v_1^2+v_2^2 && -v_2^2 && \\
 && -v_2^2 && v_2^2+v_3^2 &-v_3^2 & \\
 &&&& \ddots & \ddots & \\
 && &&&    -v_{n-1}^2 &\  v_{n-1}^2 \end{array} \right) \,\,\,.
\end{equation}
The mass matrix for the link field fermions and the gauginos is such that
\begin{equation}
M^2_{\rm fermions}=2 g^2
\left(\begin{array}{cc} \Theta \Theta^\dagger &  \\
&  \Theta^\dagger \Theta
\end{array} \right)\  \,\,\,,
\end{equation}
where the $n\times(n-1)$ dimensional matrix $\Theta$ is given by
\begin{equation}
\Theta=\left(\begin{array}{cccc}
v_1 & & & \\
-v_1 & v_2 & & \\
 & \ddots & \ddots & \\
 & & -v_{n-2} & v_{n-1} \\
 & & & -v_{n-1} \end{array} \right)\ .
\end{equation}
Clearly, $2 g^2 \Theta \Theta^\dagger \equiv m_{\rm gauge}^2$, so that half
of the fermion spectrum coincides with the gauge boson spectrum.  The scalar
spectrum, on the other hand, may be obtained by expanding Eq.~(\ref{eq:pot})
about the its minimum.  One finds
\begin{equation}
{\cal L}\supset\frac{1}{2}(\varphi_i^\dagger \ |\ \varphi_i)\,g^2\left(\begin{array}{c|c}\Theta^\dagger
\Theta \ & \ \Theta^\dagger \Theta  \\ \hline
\Theta^\dagger \Theta\  & \ \Theta^\dagger \Theta \end{array}\right)
\left(\begin{array}{c}\varphi_i
\\ \hline \varphi_i^\dagger \end{array}\right) \,\,\,.
\end{equation}
The imaginary modes $(\varphi_i-\varphi_i^\dagger)/\sqrt{2}$ have vanishing masses, and correspond
to the would-be Goldstone bosons of the spontaneous symmetry breaking U(1)$^n \rightarrow U(1)$.  The real
modes $(\varphi_i+\varphi_i^\dagger)/\sqrt{2}$ have the mass matrix
\begin{equation}
M^2_{scalars} = 2 g^2 \Theta^\dagger \Theta  \,\,\,
\end{equation}
which coincides precisely with the remaining massive fermion modes.  Finally, the $\bar{\phi}$ scalars and
their fermionic partners remain massless.  Although $n$ FI terms are present, we conclude that the KK
spectrum remains exactly supersymmetric.

This peculiar result can be understood by considering a simpler theory: a 4D ${\cal N}=1$ SUSY U(1)
gauge theory with no matter, plus an FI term.  This theory also has an exactly supersymmetric spectrum. The
sole effect of the FI term is to introduce a cosmological constant, which is irrelevant if gravity
is not included.   Precisely the same is true in our model.  One can show that the potential
Eq.~(\ref{eq:pot}) has a non-vanishing vacuum energy density $(\sum_i \xi_i)^2/n$.

The effects of SUSY breaking reappear in the particle spectrum if the model is coupled
to another sector.  Imagine that we introduce a vector-like pair of chiral superfields
that are charged only under the first U(1) factor.  The nonvanishing $D_1$ vev will split the squared
masses of their scalar components by $\pm 2\left<D_1\right>$.  If these fields are
also charged under the gauge groups of the minimal supersymmetric standard model (MSSM), then SUSY-breaking
effects will be gauge mediated to the observable sector.  Interestingly, the scale of SUSY breaking that is
relevant to gauge mediation is determined by a single D-term vev, $D_1$, while the scale relevant to
gravity-mediation is set by all $n$ non-vanishing D-terms.   Gravity-mediated SUSY-breaking effects
therefore scale with the size of the moose.  It is possible in such a model to have competing effects from the
gauge and gravity mediation of supersymmetry breaking and a heavier gravitino than in other D-term supersymmetry
breaking scenarios.

\section{Conclusions}

We have presented a supersymmetric U(1) gauge theory that deconstructs a warped extra dimension
and dynamically generates a warp factor.   The warping is accomplished via Fayet-Iliopoulos D-terms
that force the squares of the link field vevs to grow by an additive factor as one moves along the
moose.   In its simplest form, the model has the peculiar feature that supersymmetry breaking appears
only via the generation of a cosmological constant, while the spectra of the physical gauge and link
states remains supersymmetric.  In the case where the moose is allowed to couple to additional matter,
the delocalization of supersymmetry breaking implies that fields localized at a single site
experience a source of supersymmetry-breaking, $D_i^2$, that is $1/n$ as strong as
the full amount available for gravity mediation leading, for example, to a heavy gravitino.
In addition, supersymmetry breaking is supersoft~\cite{ss} in this scenario. These features may make
our U(1)$^n$ model distinctive if it is applied as a secluded supersymmetry-breaking sector for
the minimal supersymmetric standard model.



%
\bibliographystyle{aipproc}   

\begin{thebibliography}{9}

\bibitem{decon}N.~Arkani-Hamed, A.~G.~Cohen and H.~Georgi,
Phys.\ Rev.\ Lett.\  {\bf 86}, 4757 (2001);
C.~T.~Hill, S.~Pokorski and J.~Wang,
Phys.\ Rev.\ D {\bf 64}, 105005 (2001)
[arXiv:hep-th/0104035].

\bibitem{deconads}
L.~Randall, Y.~Shadmi and N.~Weiner,
JHEP {\bf 0301}, 055 (2003)
[arXiv:hep-th/0208120];
A.~Katz and Y.~Shadmi,
JHEP {\bf 0411}, 060 (2004)
[arXiv:hep-th/0409223].

\bibitem{thepaper}
C.~D.~Carone, J.~Erlich and B.~Glover,
JHEP {\bf 0510}, 042 (2005) [arXiv:hep-ph/0509002].

\bibitem{more1}
C.~D.~Carone,
Phys.\ Rev.\ D {\bf 71}, 075013 (2005)
[arXiv:hep-ph/0503069].

\bibitem{more2}
C.~Csaki, {\em et al.},
Phys.\ Rev.\ D {\bf 65}, 015003 (2002);
H.~C.~Cheng, K.~T.~Matchev and J.~Wang,
Phys.\ Lett.\ B {\bf 521}, 308 (2001);
P.~H.~Chankowski, A.~Falkowski and S.~Pokorski,
JHEP {\bf 0208}, 003 (2002);
C.~Csaki, G.~D.~Kribs and J.~Terning,
Phys.\ Rev.\ D {\bf 65}, 015004 (2002).
\bibitem{ss}
P.~J.~Fox, A.~E.~Nelson and N.~Weiner,
JHEP {\bf 0208}, 035 (2002).
\end{thebibliography}

\end{document}